\documentclass[twocolumn]{aastex6}

\shorttitle{Characterization of the Wolf 1061 Planetary System}
\shortauthors{Stephen R. Kane et al.}

\begin{document}

\title{Characterization of the Wolf 1061 Planetary System}

\author{Stephen R. Kane\altaffilmark{1},
  Kaspar von Braun\altaffilmark{2},
  Gregory W. Henry\altaffilmark{3},
  Miranda A. Waters\altaffilmark{1},
  Tabetha S. Boyajian\altaffilmark{4}
  Andrew W. Mann\altaffilmark{5}
}
\email{skane@sfsu.edu}
\altaffiltext{1}{Department of Physics \& Astronomy, San Francisco
  State University, 1600 Holloway Avenue, San Francisco, CA 94132,
  USA}
\altaffiltext{2}{Lowell Observatory, 1400 West Mars Hill Road,
  Flagstaff, AZ 86001, USA}
\altaffiltext{3}{Center of Excellence in Information Systems, Tennessee
  State University, 3500 John A. Merritt Blvd., Box 9501, Nashville,
  TN 37209, USA}
\altaffiltext{4}{Department of Physics \& Astronomy, Louisiana State
  University, Baton Rouge, LA 70803, USA}
\altaffiltext{5}{Department of Astronomy, University of Texas at
  Austin, Austin, TX 78712, USA}


\begin{abstract}

A critical component of exoplanetary studies is an exhaustive
characterization of the host star, from which the planetary properties
are frequently derived. Of particular value are the radius,
temperature, and luminosity, which are key stellar parameters for
studies of transit and habitability science. Here we present the
results of new observations of Wolf~1061, known to host three
super-Earths. Our observations from the Center for High Angular
Resolution Astronomy (CHARA) interferometric array provide a direct
stellar radius measurement of $0.3207 \pm 0.0088$~$R_{\odot}$, from
which we calculate the effective temperature and luminosity using
spectral energy distribution models. We obtained seven years of
precise, automated photometry that reveals the correct stellar
rotation period of $89.3\pm1.8$~days, finds no evidence of photometric
transits, and confirms that the radial velocity signals are not due to
stellar activity. Finally, our stellar properties are used to
calculate the extent of the Habitable Zone for the Wolf~1061 system,
for which the optimistic boundaries are 0.09--0.23~AU. Our simulations
of the planetary orbital dynamics show that the eccentricity of the
Habitable Zone planet oscillates to values as high as $\sim$0.15 as it
exchanges angular momentum with the other planets in the system.

\end{abstract}



\section{Introduction}
\label{intro}

It is frequently stated that we understand exoplanets only as well as
we understand the host star. Such a statement is particularly true for
low-mass dwarf stars, whose atmospheres often diverge from blackbody
models. There has been a concerted effort in recent years to obtain
observational constraints on the stellar models for low-mass stars
\citep{boy12,man15b}, especially for those monitored by the {\it
  Kepler} mission \citep{mui12,mui14,hub14,gai16}. A further challenge
includes the confusion that can be caused by the stellar rotation
period of low-mass stars since that can often coincide with the range
of orbital periods of planets that may exist in the Habitable Zone
(HZ) of those stars \citep{new16a,van16}. Even so, there have been
several successful detections of terrestrial planets in or near the HZ
of low-mass stars, such as Kepler-186~f \citep{qui14}, K2-3~d
\citep{cro15}, and the recently discovered Proxima Centauri b
\citep{ang16}.

The low-mass M-dwarf star, Wolf 1061 (also designated as GJ~628), is
one of our closest neighbors, located approximately 4.3 pc away
\citep{van07}. The star was recently discovered to host three planets
that lie within the super-Earth mass regime, one of which may be
located within the HZ of the system \citep{wri16}. The orbits of the
planets were significantly updated by \citet{ast16b} using additional
data from the High Accuracy Radial velocity Planet Searcher (HARPS)
spectrograph. The orbital solutions largely agree with respect to the
inner two planets, but the \citet{ast16b} solution finds an orbital
period of the outer planet that is a factor of $\sim$3 larger than
that found by \citet{wri16}. Both of the solutions do agree that
planet c is near or in the HZ of the host star, the location of which
is highly dependent on the star parameters of luminosity and effective
temperature. Both solutions also predict reasonably high transit
probabilities and depths such that follow-up photometry during
calculated transit windows is encouraged.

This paper presents a characterization of the Wolf~1061 host star and
the associated planets with new data and numerical simulations. In
Section~\ref{fund}, the fundamental stellar parameters of the host
star are updated through interferometry data that provide measurements
of the stellar radius, effective temperature, and
luminosity. Section~\ref{phot} presents seven years of precise,
automated photometry that reveals the correct stellar rotation period,
supports the existence of the three purported planets, and finds no
evidence for planetary transits. The revised stellar parameters are
utilized in Section~\ref{hab} with a calculation of the system HZ and
a dynamical simulation that shows variation of eccentricities for the
planetary orbits. We provide a concluding discussion of the results in
Section~\ref{conclusions}.


\section{Fundamental Stellar Parameters}
\label{fund}

Wolf~1061 has been an object of interest for quite some time,
primarily because of its high proper motion of $1191.5\pm0.9$~mas/yr
\citep{dav15} and its membership in the solar neighborhood. Despite
its relatively close position to our star, the stellar properties of
Wolf~1061 have remained uncertain, largely due to the fact that it is
a very dim M3V late-type dwarf. Prior to our investigation, the values
of Wolf 1061's temperature, luminosity, flux, and radius have varied
in the literature. Previous estimates for the red dwarf's temperature
have been reported as being as low as 2877~K \citep{leg15} to as high
as 3400~K \citep{ave12} with a variety of values reported between
these two extremes
\citep{jen09,bai12,one12,roj12,can13,nev14,lin16,wri16}. Less extreme
are the variations in radius, from 0.30~$R_\odot$ \citep{new16a} to
$0.325\pm0.012$~$R_\odot$ \cite{man15b}, and in luminosity, which was
previously reported to be 0.007870~$L_\odot$ \citep{wri16}. Precise
determination of these essential characteristics have taken on new
importance in light of the early 2016 discovery that Wolf~1061 is host
to three exoplanets \citep{wri16,ast16b}.


\subsection{Stellar Radius}
\label{sec:stellar_radius}

Wolf~1061 was observed on June 30 and August 3--4, 2016, using the
Georgia State University Center for High Angular Resolution Astronomy
(CHARA) interferometric array \citep{ten05}. Observations were
conducted in $H$-band with the CHARA Classic beam combiner
\citep{stu03,ten05} in single-baseline mode. To remove the influence
of atmospheric and instrumental systematics, our interferometric
observations consist of bracketed sequences of object and calibrator
stars. Calibrators were chosen using the {\tt ASPRO} tool\footnote{\tt
  http://www.jmmc.fr/aspro} to be near-point-like sources of similar
brightness as Wolf~1061, located at small angular distances from it,
and observed directly before and after the target: HD~143459,
HD~146254, HD~149013, and HD~153229. This procedure follows our
requirements that we use at least two calibrators, two baselines, and
data obtained during at least two nights \citep[e.g.,][and references
  therein]{von14,boy15}.

\begin{figure}
  \includegraphics[width=8.2cm]{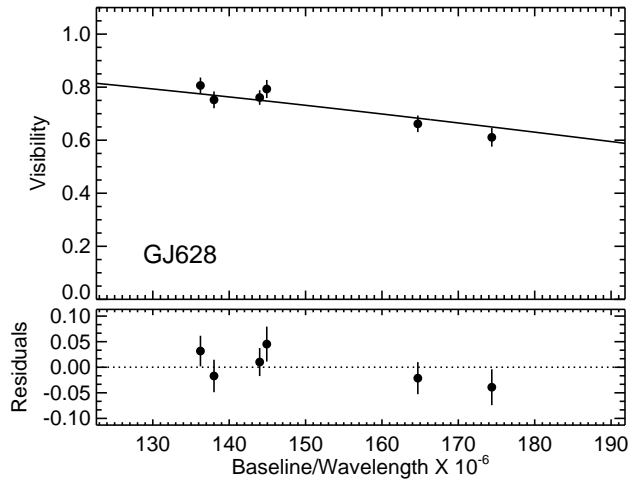}
  \caption{Calibrated visibility observations along with the
    limb-darkened angular diameter fit for Wolf~1061 (GJ~628) (top
    panel) along with the fractional residuals around the fit (bottom
    panel). For more details, see \S \ref{sec:stellar_radius}.}
  \label{fig:diameters}
\end{figure}

The uniform-disk and limb-darkened angular diameters ($\theta_{\rm             
  UD}$ and $\theta_{\rm LD}$, respectively) are calculated by fitting
the calibrated visibility measurements (Figure \ref{fig:diameters}) to
the respective functions for each relation. These functions may be
described as $n^{th}$-order Bessel functions of the angular diameter
of the star, the projected distance between the two telescopes, and
the wavelength of observation \citep{han74}.

We use the linear limb-darkening coefficient $\mu_{H} = 0.376$ from
the PHOENIX models in \citet{cla11} for stellar $T_\mathrm{eff}$ =
3000 K and $\log g$ = 4.5 to convert from $\theta_{\rm UD}$ to
$\theta_{\rm LD}$. The uncertainties in the adopted limb-darkening
coefficient amount to 0.2\% when modifying the adopted gravity by
0.5~dex or the adopted $T_\mathrm{eff}$ by 200K, well within the
errors of our diameter estimate.

Our interferometric measurements produce the following values for
Wolf~1061: $\theta_{\rm UD} = 0.674 \pm 0.018$~milliarcseconds (mas)
and $\theta_{\rm LD} = 0.695 \pm 0.018$~mas. Combined with the
trigonometric parallax measurement of $232.98\pm1.60$~mas from
\citet{van07}, we obtain a stellar radius for Wolf~1061 of $0.3207 \pm
0.0088$~$R_{\odot}$, which is practically identical to the one
estimated in \citet{man15b} of $0.325 \pm 0.012$~$R_{\odot}$.


\subsection{Stellar Effective Temperature and Luminosity}
\label{sec:effective_temperature}

To calculate Wolf~1061's effective temperature and luminosity, we
perform a spectral energy distribution (SED) fit based on
spectrophotometry data obtained as part of the survey described in
\citet{man15b}; see in particular their Section 3. These
spectrophotometry data have no color terms and only require a
zero-point offset. We use literature photometry from \citet{joh54,
  nic57, joh65, cor72, vee74, mou76, cou80a, cou80b, rei82, wei84,
  mer86, wei86, wei87, bei88, lai89, bes90, wei96, koe02, cut03,
  gau07, kil07, koe10, hen12, tur15, wri16} to scale the
spectrophotometry data and obtain the bolometric flux by simply
integrating over wavelength. Interstellar reddening is set to zero in
the fit, due to the close proximity of Wolf~1061. In the calculation
of the bolometric flux, we use the modified filter profiles for the
literature photometry from \citet{man15a} and use the 2\% error
correction described by \citet{boh14} to obtain realistic error
estimates in $F_{\rm BOL}$. We calculate the following for Wolf~1061:
$F_{\rm BOL} = (1.920 \pm 0.043)\times10^{-8}$~erg~cm$^{-2}$ s$^{-1}$,
$L = 0.01102 \pm 0.00027$~$L_{\odot}$, and $T_{\rm eff} = 3305 \pm
46$~K. These values are consistent at $\lesssim 1 \sigma$ with the
ones in Table 5 of \citet{man15b} that use interferometric data for
calibration of their semi-empirical methods. Our stellar parameters
for Wolf~1061, including the rotation period described in
Section~\ref{rot}, are summarized in Table~\ref{stellar}.

\begin{deluxetable}{lc}
  \tablecaption{\label{stellar} Stellar Parameters}
  \tablehead{
    \colhead{Parameter} &
    \colhead{Value}
  }
  \startdata
  $V$                                       & 10.07 \\
  $B-V$                                     & 1.57 \\
  Distance (pc)                             & $4.29 \pm 0.03$ \\
  $F_\mathrm{BOL}$ (erg~cm$^{-2}$~s$^{-1}$) & $(1.920\pm0.043)\times10^{-8}$ \\
  $T_\mathrm{eff}$ (K)                      & $3305 \pm 46$ \\
  $R_\star$ ($R_\odot$)                     & $0.3207 \pm 0.0088$ \\
  $L_\star$ ($L_\odot$)                     & $0.01102 \pm 0.00027$ \\
  $P_\mathrm{rot}$ (days)                   & $89.3\pm1.8$ \\
  \enddata
\end{deluxetable}


\section{Photometric Observations}
\label{phot}

\begin{figure}
  \includegraphics[width=8.2cm]{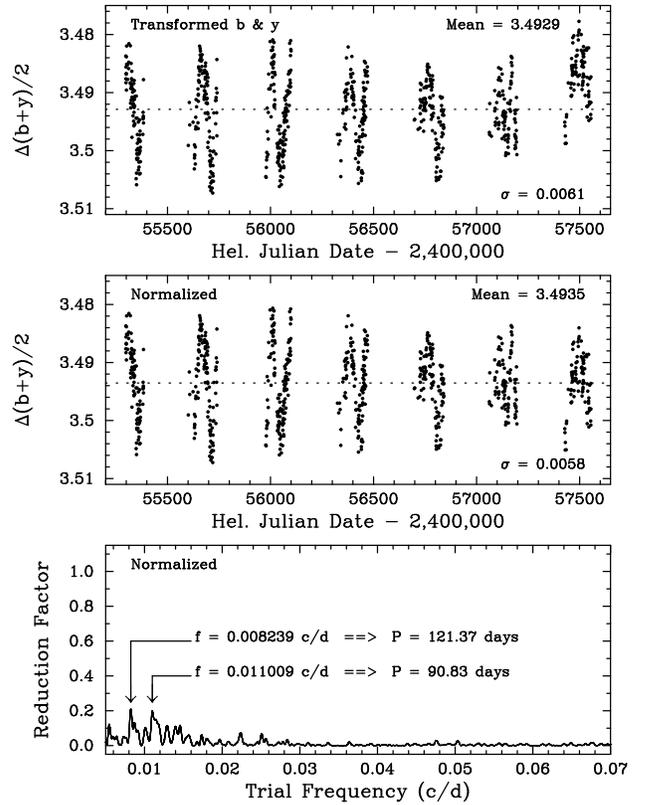}
  \caption{$Top$: Seven years of photometric observations of
    Wolf~1061, comprising 756 nightly measurements, acquired with the
    T11 0.8~m APT at Fairborn Observatory. Slow rotational modulation
    of dark spots on the star's surface, as well as year-to-year
    evolution of the spot distribution, account for the brightness
    variability. The dotted line marks the mean brightness.  $Middle$:
    The 756 observations are normalized so that all seasonal means are
    equal to the first, marked by the dotted line.  $Bottom$:
    Frequency spectrum of the complete normalized data set revealing
    low-amplitude variability at 121 or 91 days.}
  \label{photfig1}
\end{figure}

We have observed Wolf~1061 during its past seven observing seasons
with the Tennessee State University (TSU) T11 0.80~m automatic
photoelectric telescope (APT) at Fairborn Observatory in Arizona.
Between 2010 April and 2016 June, the APT acquired 756 brightness
measurements of Wolf~1061 on 464 different nights.  Like other TSU
APTs, T11 is equipped with a two-channel precision photometer designed
and built by Louis Boyd at Fairborn.  The photometer uses a dichroic
filter and two EMI 9124QB bi-alkali photomultiplier tubes to separate
and simultaneously measure the Str\"omgren $b$ and $y$ photometric
passbands. Wolf~1061, designated here as the program star (P,
$V=10.10$, $B-V=1.60$, M3.5~V), was observed differentially with
respect to three constant comparison stars HD~150177 (C1, $V=6.33$,
$B-V=0.49$, F3~V), HD~147753 (C2, $V=7.58$, $B-V=0.55$, F2~V), and
HD~148968 (C3, $V=6.98$, $B-V=0.14$, A0~V).  All differential
magnitudes were corrected for extinction and transformed to the
Str\"omgren photometric system.  We computed final differential
magnitudes of Wolf~1061 against the mean brightness of all three
comparison stars as $P-(C1+C2+C3)/3_{by}$, where the subscript $by$
indicates that we combined the Str\"omgren $b$ and $y$ observations
into a single $(b+y)/2$ passband.  The precision of a single
observation from the T11 APT is typically $0.0015-0.0020$~mag,
determined by intercomparison of the comparison stars. Further details
of our automatic telescopes, precision photometers, and observing and
data reduction procedures can be found in \citet{hen99} and
\citet{eat03} and references therein. Note that the T11 APT is
essentially identical to the T8 APT described in \citet{hen99}.


\subsection{Stellar Rotation Period}
\label{rot}

The final $P-(C1+C2+C3)/3_{by}$ differential magnitudes are plotted in
the top panel of Figure~\ref{photfig1} and show Wolf~1061 to be
varying over a range of $\sim0.02$~mag with a timescale of
$\sim100$~days. To remove the effects of the small year-to-year
variations in mean brightness, we normalized the data shown in the
middle panel of Figure~\ref{photfig1} by adjusting the seven seasons
so that each has the same mean as the first. A frequency spectrum of
the normalized data is shown in the bottom panel. Weak periodicity is
found around 91 and 121 days, both with amplitudes of a few
millimags. The low amplitudes are due not only to the intrinsically
low amplitude of Wolf~1061's photometric variability but also to
year-to-year changes in the amplitude, shape, mean magnitude, and
phase of minimum of the light curve.  The 91 and 121 day periods are
yearly aliases of each other caused by the large seasonal gaps in our
light curve.  We take the variability in Wolf~1061 to arise from the
rotational modulation of a slowly evolving spot distribution on the
photosphere of the star.

\begin{deluxetable*}{ccccccc}
  \tabletypesize{\small}
  \tablecolumns{7}
  \tablewidth{0pc}
  \tablecaption{\label{phottab} Summary of Photometric Observations for Wolf~1061}
  \tablehead{
    \colhead{Observing} & \colhead{} & \colhead{Julian Date Range} &
    \colhead{Sigma} & \colhead{$P_{rot}$} & \colhead{Full Amplitude} &
    \colhead{$<P-(C1+C2+C3)/3_{by}>$} \\
    \colhead{Season} & \colhead{$N_{obs}$} & \colhead{(HJD $-$ 2,400,000)} &
    \colhead{(mag)} & \colhead{(days)} & \colhead{(mag)} &
    \colhead{(mag)} \\
    \colhead{(1)} & \colhead{(2)} & \colhead{(3)} &
    \colhead{(4)} & \colhead{(5)} & \colhead{(6)} &
    \colhead{(7)}
  }
  \startdata
  2010 &  92 & 55297--55383 & 0.0061 &   (103.9)    &      (0.015)      & $3.4935 \pm 0.0006$ \\
  2011 & 117 & 55601--55740 & 0.0070 & $92.9\pm3.4$ & $0.0186\pm0.0009$ & $3.4936 \pm 0.0006$ \\
  2012 & 136 & 55976--56099 & 0.0068 & $84.2\pm3.0$ & $0.0166\pm0.0008$ & $3.4938 \pm 0.0006$ \\
  2013 & 115 & 56325--56470 & 0.0055 & $91.4\pm2.9$ & $0.0147\pm0.0009$ & $3.4937 \pm 0.0005$ \\
  2014 & 111 & 56698--56840 & 0.0050 &   (108.3)    &      (0.011)      & $3.4937 \pm 0.0005$ \\
  2015 &  93 & 57062--57195 & 0.0042 &    (58.7)    &      (0.008)      & $3.4936 \pm 0.0004$ \\
  2016 &  92 & 57428--57554 & 0.0045 &   (199.2)    &      (0.015)      & $3.4872 \pm 0.0005$ \\
  \enddata
  \tablecomments{Periods and Amplitudes in parentheses are poorly determined.}
\end{deluxetable*}

\begin{figure}
  \includegraphics[width=8.2cm]{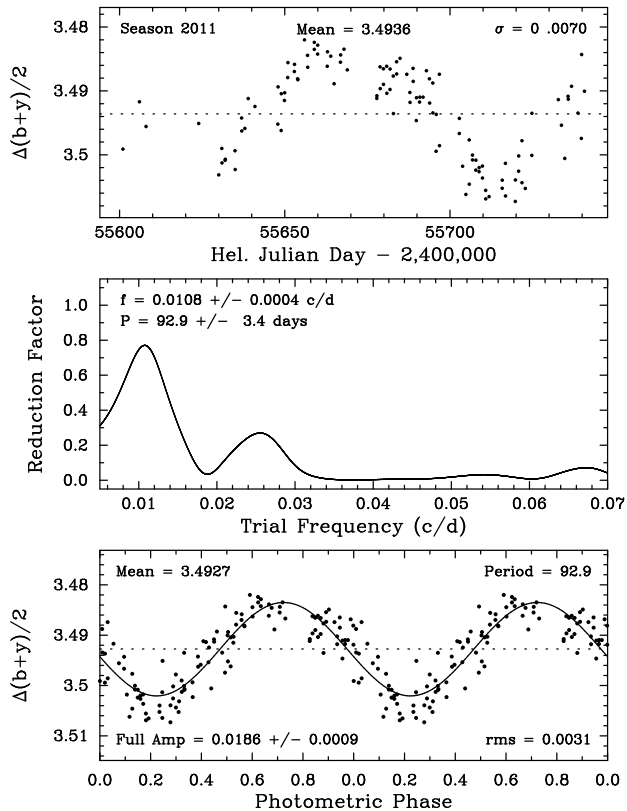}
  \caption{$Top$: JD plot of the 2011 observing season of Wolf~1061.
    $Middle$: Frequency spectrum of the 2011 data giving a photometric
    period of $92.9\pm3.4$~days. $Bottom$: The 2011 data phased to the
    best period of 92.9~days and showing a peak-to-peak amplitude of
    0.019~mag.}
  \label{photfig2}
\end{figure}

\begin{figure}
  \includegraphics[width=8.2cm]{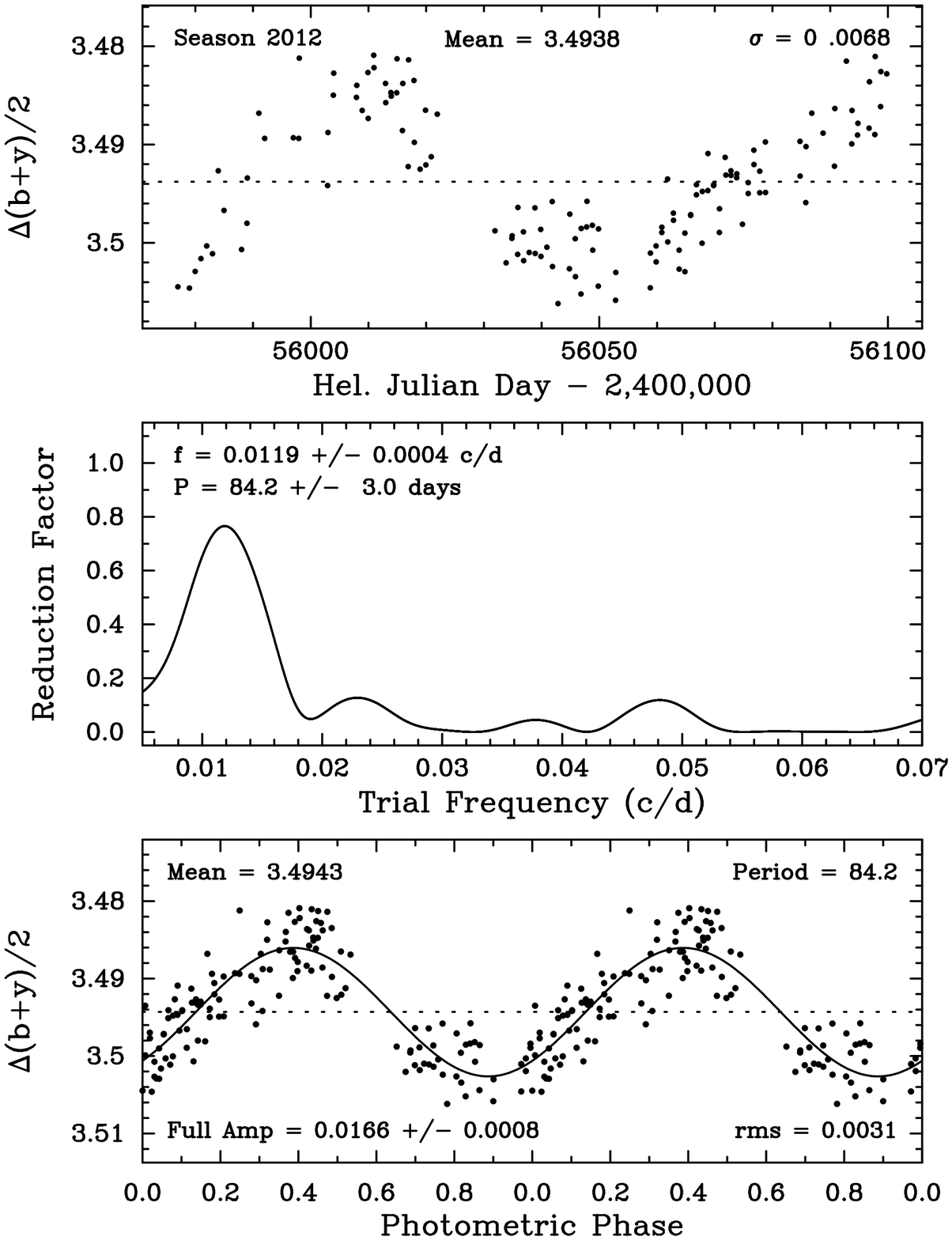}
  \caption{Same as Figure~\ref{photfig2} except for the 2012 observing
    season, giving a period of $84.2\pm3.0$~days and a peak-to-peak
    amplitude of 0.017~mag.}
  \label{photfig3}
\end{figure}

\begin{figure}
  \includegraphics[width=8.2cm]{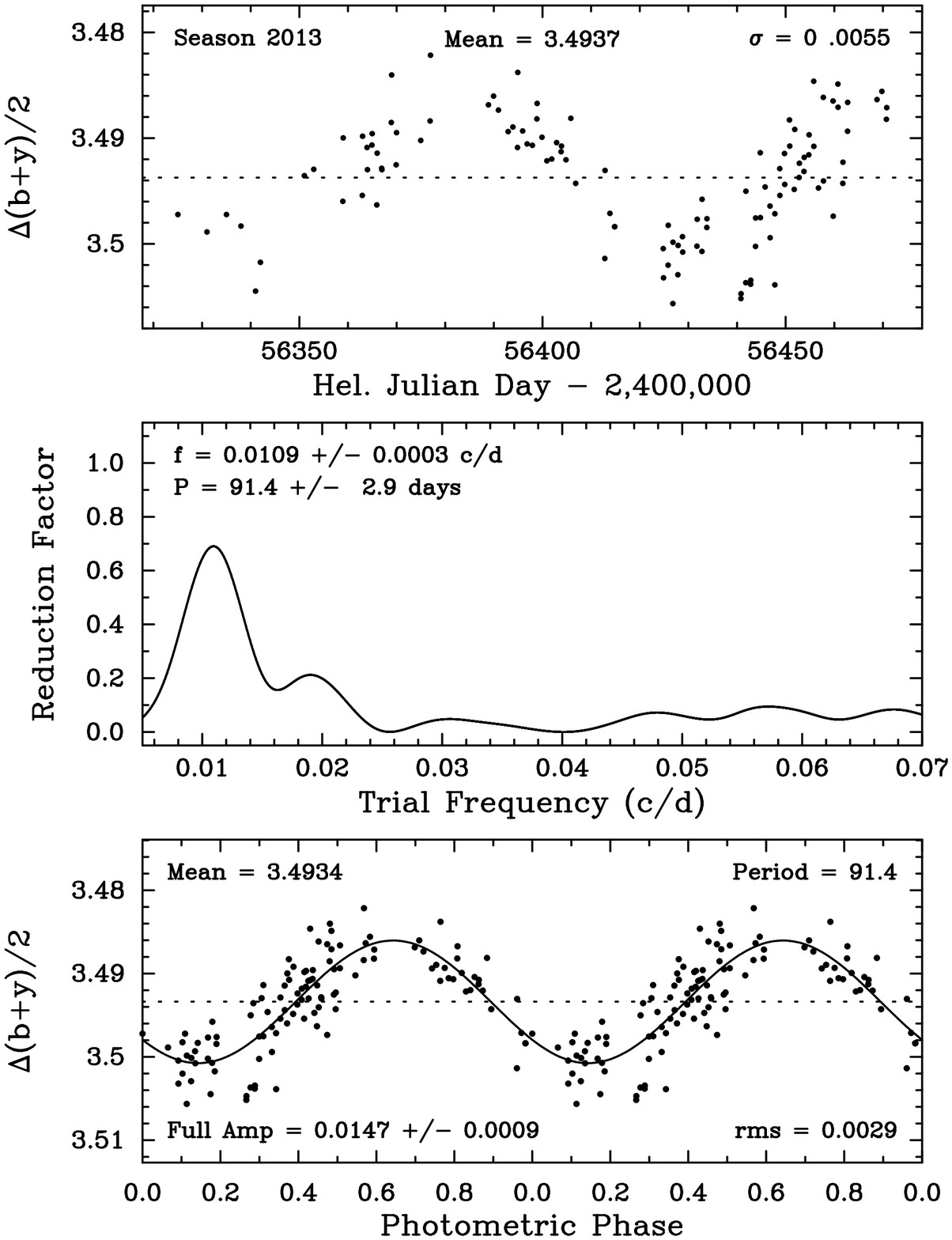}
  \caption{Same as Figure~\ref{photfig2} except for the 2013 observing
    season, giving a best period of $91.4\pm2.9$~days and a
    peak-to-peak amplitude of 0.015~mag.}
  \label{photfig4}
\end{figure}

To determine the correct rotation period, we attempted a periodogram
analysis of the seven individual observing seasons.  Only the 2011,
2012, and 2013 observing seasons cover the light curve sufficiently
well to give reliable results.  Frequency spectra for these three
observing seasons are shown in Figures~\ref{photfig2}, \ref{photfig3},
\& \ref{photfig4}. The complete results of our seasonal photometric
analysis of Wolf~1061 are given in Table~\ref{phottab}. Rotation
periods and amplitudes that are poorly constrained are given in
parentheses. The weighted mean of the 2011, 2012, and 2013 photometric
periods is $89.3\pm1.8$~days. We identify this period as the true
stellar rotation period and the 121-day period from
Figure~\ref{photfig1} as its yearly alias. This is consistent with the
93~day rotation period determined for Wolf~1061 by \citet{ast16a} from
analyses of the Ca II H\&K emission lines. Using the kinematic work of
\citet{new16b} for nearby M dwarfs, the rotation period suggests an
age for Wolf~1061 of $>5$~Gyrs.

However, as seen above in Figure~\ref{photfig1}, the rotational
modulation of the light curve is not strictly sinusoidal over the seven
year span of our observations. Therefore, stellar activity on the
surface of Wolf~1061 might still be responsible for radial velocity
variability, as has been demonstrated in other moderately active stars
\citep[see, e.g.,][]{que01,pau04,boi12}.


\subsection{Ruling Out Planetary Transits}
\label{tran}

\begin{figure}
  \includegraphics[width=8.2cm]{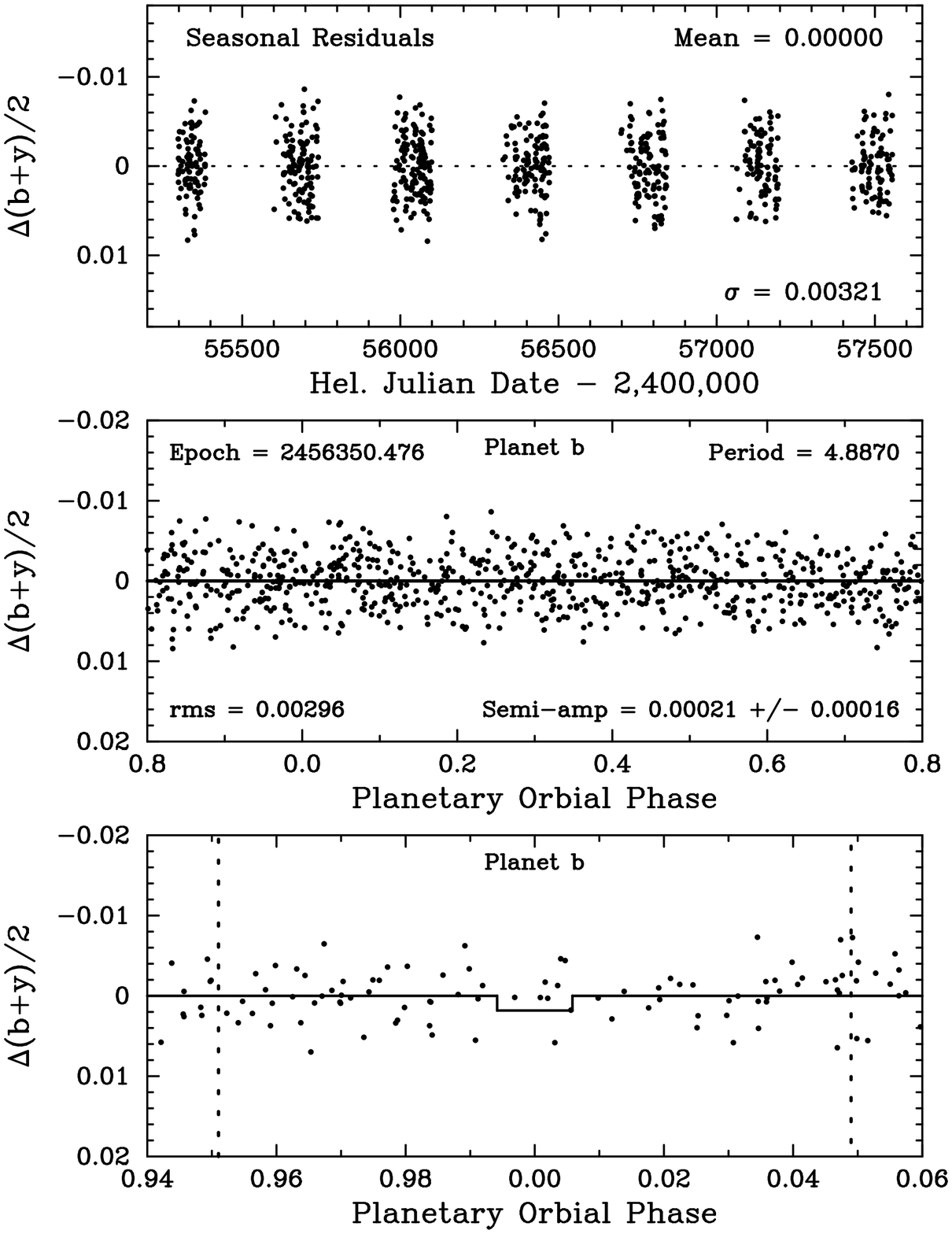}
  \caption{$Top$: Residuals from the individual sine fits to the seven
    observing seasons of Wolf~1061 summarized in Table~\ref{phottab},
    plotted against Julian Date. $Middle$: Residuals from the top
    panel phased with the 4.8870-day orbital period of planet~b and
    time of conjunction derived from the radial velocities.  A
    least-squares sine fit on the radial velocity period gives a
    semi-amplitude of just $0.00021\pm0.00016$~mag, establishing to
    high precision the lack of stellar activity on the radial velocity
    period and thus confirming the presence of stellar reflex motion
    caused by an orbiting planet. $Bottom$: Close-up of the
    observations near the time of planetary conjunction at phase
    0.0. The solid line shows a model transit computed from the
    parameters of planet~b.  The vertical lines mark the uncertainty
    in the predicted transit window. Our current photometric
    observations provide no evidence for transits.}
  \label{photfig5}
\end{figure}

\begin{figure}
  \includegraphics[width=8.2cm]{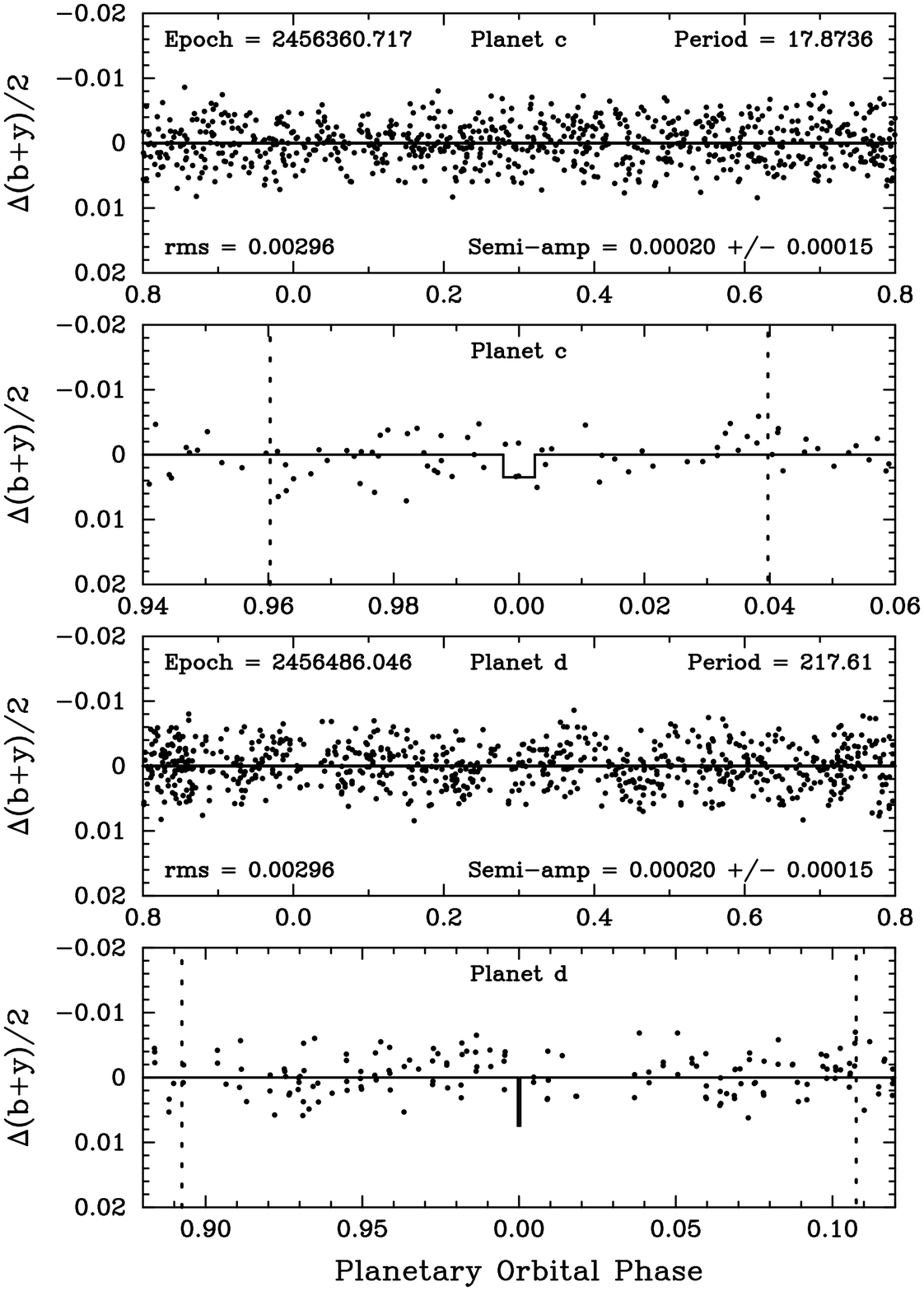}
  \caption{Same as Figure~\ref{photfig5} but for Wolf~1061 c and d.}
  \label{photfig6}
\end{figure}

We remove most of the photometric variability in Wolf~1061 by taking
the residuals from the yearly sine fits specified in
Table~\ref{phottab}. These are plotted in the top panel of
Figure~\ref{photfig5}. The standard deviation of the residuals from
the mean (marked by the dotted line) is 0.0032~mag, roughly half the
variability of the original light curve in the top panel of
Figure~\ref{photfig1}. Periodogram analysis of the full set of
residuals finds no significant periodicity. The middle panel of
Figure~\ref{photfig5}, shows the seasonal residuals phased with the
4.8870-day orbital period of Wolf~1061~b and the epoch of mid-transit
from Table~4 of \citet{ast16b}. A least-squares sinusoidal fit to the
phased data gives a formal semi-amplitude of just
$0.00021~\pm~0.00016$ mag, which limits any periodic brightness
variability on the orbital period to a very small fraction of one
milli-magnitude (mmag). This rules out the possibility that the
4.8870-day radial velocity variations are due to stellar
activity. Instead, the lack of photometric variability confirms that
the radial velocity variations at 4.8870~day result from true
planetary reflex motion.

The photometric observations within $\pm0.06$ orbital phase of
mid-transit are plotted with an expanded scale in the bottom panel of
Figure~\ref{photfig5}. The solid curve shows the predicted transit
phase (0.0), depth ($0.00183$ mag), and duration ($0.057$~days) of a
central transit, computed from the stellar and planetary radii and the
orbital elements of Wolf~1061~b. The vertical dotted lines give the
$\pm1\sigma$ uncertainty in the timing of the transit window, based on
the uncertainties in the stellar radius provided in
Section~\ref{sec:stellar_radius} and the improved orbital elements
from \citet{ast16b}. We find no evidence in our data for transits of
planet~b.

Results of similar analyses for planets~c and d are shown in
Figure~\ref{photfig6}. The low photometric amplitude of
$0.00020~\pm~0.00015$~mag in the top panel confirms Wolf~1061~c as a
planet since it shows that the radial velocity variations are not due
to stellar activity. In the second panel, we find no evidence of
photometric transits of planet~c. The low amplitude of
$0.00020~\pm~0.00015$~mag in the third panel confirms Wolf~1061~d as a
planet.  The bottom panel shows we have insufficient data to rule out
transits of planet~d.


\section{Habitability of the System}
\label{hab}

Using the stellar parameters described above, we calculate the
boundaries of the HZ and the long-term stability of the planets with
respect to the HZ.


\subsection{The Habitable Zone}
\label{hz}

The concept of the HZ as a target selection tool has been developed
based on Earth climate models for several decades
\citep{kas93,kas14}. Specifically, the HZ defines the locations around
a host star where the climate of an Earth analog will remain cool
enough to avoid a runaway greenhouse effect and warm enough to prevent
a runaway snowball effect. These calculations account for water
absorption in the planetary atmosphere and stellar properties such as
luminosity and effective temperature. To calculate the extent of the
HZ, we use the methodology of \citet{kop13,kop14}. The
``conservative'' and ``optimistic'' HZ boundaries are calculated based
on assumption regarding the time span over which the atmospheric
evolutionary history of Venus and Mars allowed liquid water to remain
on the surface. The catalog of confirmed planets and planetary
candidates detected by the {\it Kepler} mission \citep{kan16}
describes the conservative and optimistic HZ boundaries in more
detail.

\begin{figure}
  \includegraphics[angle=270,width=8.5cm]{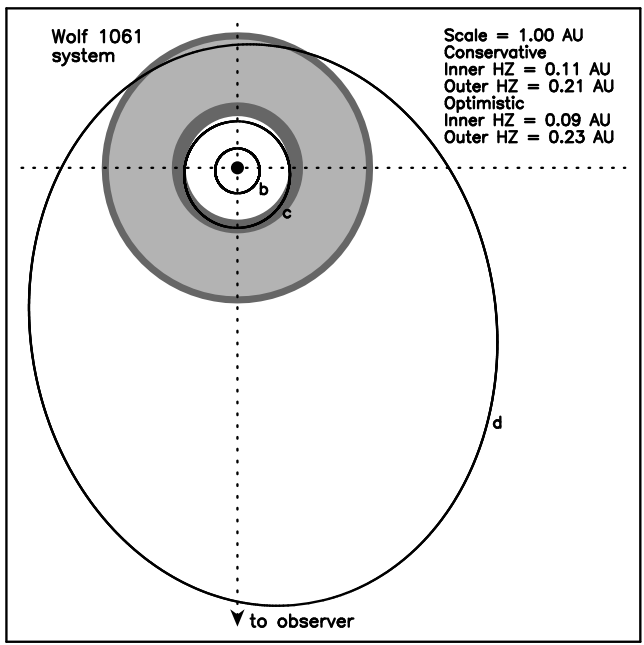}
  \caption{A top-down view of the Wolf~1061 system showing the orbits
    of the planets overlaid on the HZ. The extent of the HZ was
    calculated using the stellar parameters from
    Section~\ref{sec:effective_temperature}. The physical scale
    depicted is 1.0~AU on a side. The conservative HZ is shown as
    light-gray and the optimistic extension to the HZ is shown as
    dark-gray.}
  \label{hzfig}
\end{figure}

Using the updated stellar parameters from this work (see
Section~\ref{sec:effective_temperature}), we estimate the inner and
outer boundaries of the conservative HZ to be 0.11~AU and 0.21~AU
respectively. Allowing for the optimistic conditions for surface liquid
water, the inner and outer HZ boundaries are 0.09~AU and 0.23~AU
respectively. Shown in Figure~\ref{hzfig} is a top-down view of the
planetary orbits in the Wolf~1061 system, using the orbital solution
of \citet{ast16b}. The conservative HZ is shown as light-gray and the
optimistic extension to the HZ is shown as dark-gray. The scale of the
figure is 1.0~AU on a side. The orbital eccentricities of the planets
are 0.13, 0.13, and 0.57 for the b, c, and d planets respectively. The
outer planet passes briefly through the HZ during its periastron
passage, spending 6\% of the orbital period within the HZ. Planet c
spends 61\% of its orbit duration within the HZ, but that time remains
constrained to the optimistic HZ. In that respect, planet c is quite
similar to the case of Kepler-69~c, which was proposed to be a strong
super-Venus candidate by \citet{kan13}. Indeed, both of the inner two
planets, terrestrial in nature according to the results of both
\citet{wri16} and \citet{ast16b}, lie within the Venus Zone of the
host star \citep{kan14} and are thus possible runaway greenhouse
candidates.


\subsection{Orbital Stability and Dynamics}
\label{stab}

Another factor that plays a role in the habitability of the system is
the orbital dynamics between the planets as a function of time. To
investigate orbital stability and dynamics, we utilized the Mercury
Integrator Package \citep{cha99}, with the hybrid
symplectic/Bulirsch-Stoer integrator and a Jacobi coordinate
system. The initial conditions were set using the orbital solution of
\citet{ast16b} and the integration executed for a simulated duration
of $10^7$ years. The time resolution was set to 0.1~days in order to
adequately meet the recommended minimum time step criterion of $1/20$
of the shortest system orbital period \citep{dun98}. The orbital
architecture of the system was output in 100 year intervals.

\begin{figure*}
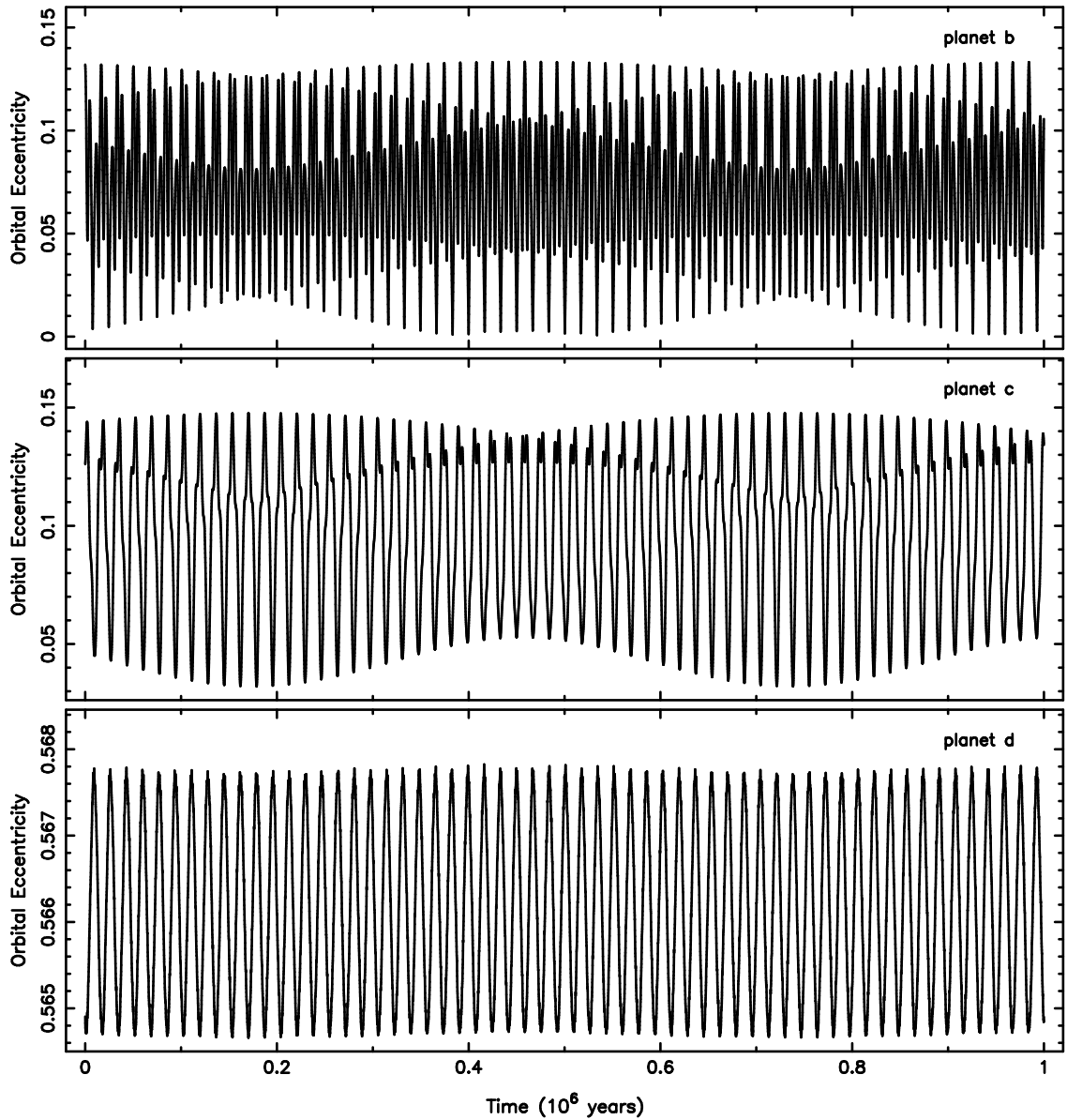

  \begin{center}
    \includegraphics[angle=270,width=15.0cm]{f09a.ps} \\
    \includegraphics[angle=270,width=15.0cm]{f09b.ps} \\
    \includegraphics[angle=270,width=15.0cm]{f09c.ps}
  \end{center}
  \caption{The eccentricity component of the orbital dynamics within
    the Wolf~1061 system, shown for planets b, c, and d (top, middle,
    and bottom panels respectively).}
  \label{ecc}
\end{figure*}

\begin{figure*}
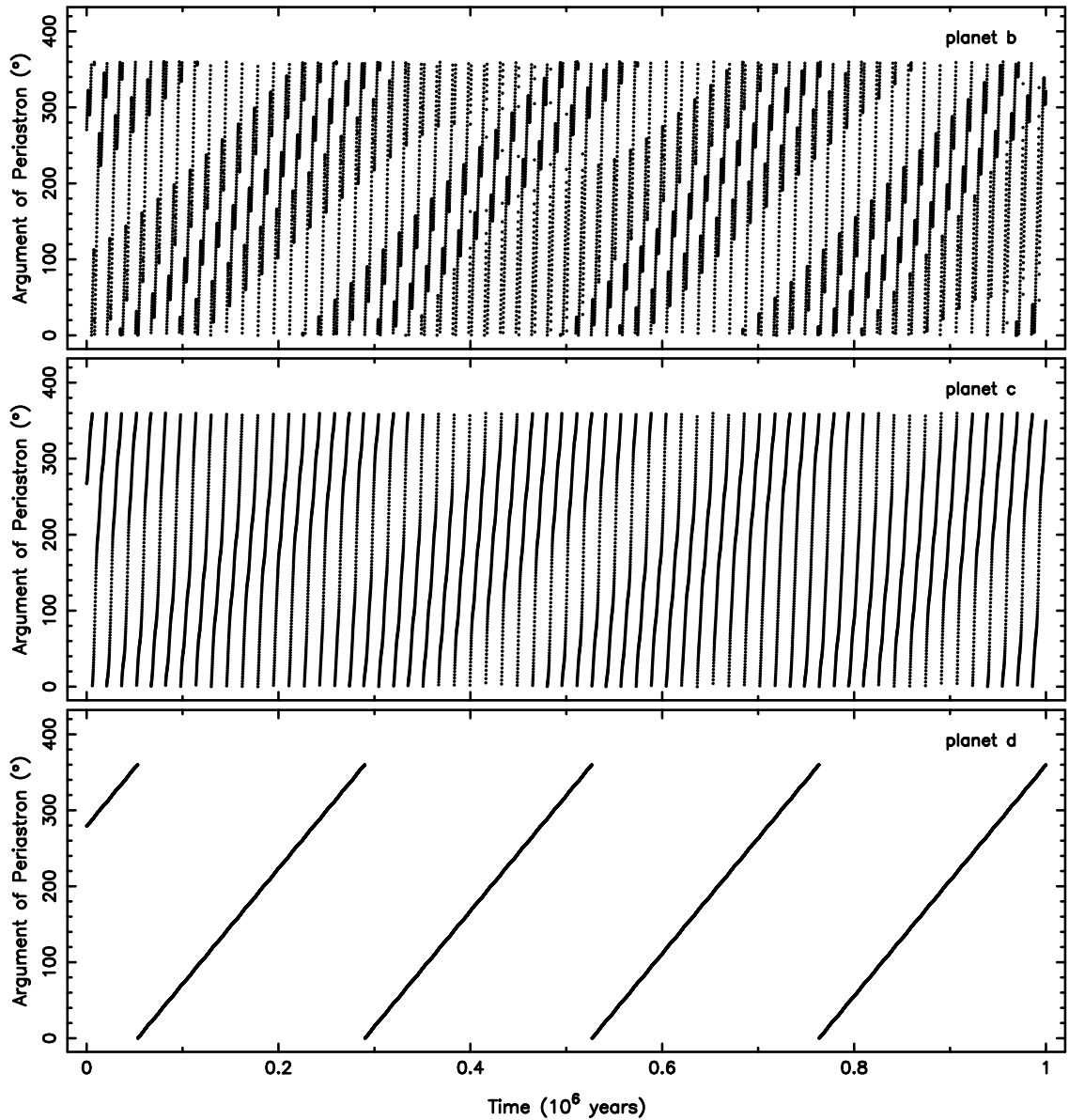

  \begin{center}
    \includegraphics[angle=270,width=15.0cm]{f10a.ps} \\
    \includegraphics[angle=270,width=15.0cm]{f10b.ps} \\
    \includegraphics[angle=270,width=15.0cm]{f10c.ps}
  \end{center}
  \caption{The argument of periastron component of the orbital
    dynamics within the Wolf~1061 system, shown for planets b, c, and
    d (top, middle, and bottom panels respectively).}
  \label{omega}
\end{figure*}

For the coplanar scenario where the system is viewed approximately
edge-on (inclination of $i = 90\degr$ and the true planetary masses
are equivalent to the minimum masses), the system was found to be
stable over the full $10^7$ year simulation duration. Although stable,
the compact nature of the system, combined with the relatively large
orbital eccentricities, results in an active dynamical evolution of
key Keplerian orbital parameters. In particular, evidence of the
angular momentum exchange between the planets can be observed in the
oscillations of the eccentricity and argument of periastron. These
evolutions of eccentricity and argument of periastron are shown for a
simulation duration of $10^6$ years in Figures \ref{ecc} and
\ref{omega} respectively.

The bottom panel of Figure~\ref{ecc} shows that the amplitude of the
eccentricity variations for the outer planet is largely insensitive to
the presence of the inner planets and remains close to the initial
value of $e_d = 0.565$. The interaction between the inner two planets
is more pronounced, with mean eccentricities substantially below the
initial values of $e_b = 0.132$ and $e_c = 0.126$. Similarly, the
precession of the periastron arguments evolves on a rapid timescale
for the inner two planets compared with the outer planet (see
Figure~\ref{omega}). The eccentricity of the inner planet reduces to
the circular case at regular intervals, whereas the eccentricity of
planet c drops as low as $\sim$0.03. The eccentricity of a planet
within the HZ does not necessarily exclude the presence of liquid
water on the surface, as the required conditions also depend on such
factors as atmospheric composition, scale height, and response to
variations in incident flux \citep{kan12,wil02}. However, it is worth
noting that a zero eccentricity for planet c results in the orbit
being entirely interior to the optimistic HZ. It is thus possible that
planet c is more amenable to habitable conditions when near to peak
eccentricity, since the planet moves slowly through the apastron
passage in the HZ.


\section{Conclusions}
\label{conclusions}

The assessment of host star properties is a critical component of
exoplanetary studies, at least for the realm of indirect detections
through which exoplanet discoveries thus far have predominantly
occurred. This situation will remain true for the coming years during
which the transit method will primarily be used from space missions
such as the Transiting Exoplanet Survey Satellite (TESS), the
CHaracterising ExOPlanet Satellite (CHEOPS), and the PLAnetary
Transits and Oscillations of stars (PLATO) mission. Of particular
interest are the radius and effective temperature of the stars since
the radius impacts the interpretation of observed transit events and
the combination of radius and temperature is used to calculate the
extent of the HZ.

Here we have presented the results from direct measurements of stellar
properties for one of the closest known exoplanet host stars,
Wolf~1061. Our direct measurement of the stellar radius from
interferometric observations gives $0.3207 \pm 0.0088$~$R_{\odot}$,
which is remarkably close to the value previous calculated by
\citet{man15b}, which can be considered a significant triumph for the
empirical calibrations used in that work. Our SED fit resulted in
determining a luminosity of $L = 0.01102 \pm 0.00027$~$L_{\odot}$ for
Wolf~1061 and, after combination with the measured angular diameter,
an effective temperature of $T_{\rm eff} = 3305 \pm 46$~K.

We further provide seven years of Wolf~1061 photometry based on
observations acquired with TSU's T11 APT. These data were sufficient
to investigate periodic signals that are a measurement of the stellar
rotation period. Our analysis was able to disentangle the various
aliases and isolate a rotation period of $89.3\pm1.8$~days. Our
photometric precision and observing cadence are able to rule out
transits of the two inner planets in the system, but the possibility of
a transiting outer planet remains open.

Finally, our measured stellar parameters were used to derive the HZ
boundaries of the system and investigate the location and dynamics of
the planetary orbits with respect to the HZ. We find that, although
the eccentric solution for planet c allows it to enter the optimistic
HZ, the two inner planets are consistent with possible super-Venus
planets \citep{kan13,kan14}. Long-term stability analysis shows that
the system is stable in the current configuration, and that the
eccentricity of the two inner planets frequently reduces to zero, at
which times the orbit of planet c is entirely interior to the
optimistic HZ. We thus conclude that the system is unlikely to host
planets with surface liquid water.


\section*{Acknowledgements}

G.W.H. acknowledges support from NASA, NSF, Tennessee State
University, and the State of Tennessee through its Centers of
Excellence program. T.S.B. acknowledges support from NASA grant
14-K2GO2\_2-0075 and Louisiana State University. This work is based
upon observations obtained with the Georgia State University Center
for High Angular Resolution Astronomy Array at Mount Wilson
Observatory. The CHARA Array is supported by the National Science
Foundation under Grant No. AST-1211929. Institutional support has been
provided from the GSU College of Arts and Sciences and the GSU Office
of the Vice President for Research and Economic Development. This
research has made use of the Jean-Marie Mariotti Center \texttt{Aspro}
service, available at www.jmmc.fr/aspro. This research has also made
use of the following archives: the Exoplanet Orbit Database and the
Exoplanet Data Explorer at exoplanets.org, the Habitable Zone Gallery
at hzgallery.org, and the NASA Exoplanet Archive, which is operated by
the California Institute of Technology, under contract with the
National Aeronautics and Space Administration under the Exoplanet
Exploration Program. The results reported herein benefited from
collaborations and/or information exchange within NASA's Nexus for
Exoplanet System Science (NExSS) research coordination network
sponsored by NASA's Science Mission Directorate.


\software{\tt ASPRO (http://www.jmmc.fr/aspro)}


\end{document}